\documentclass[preprint,12pt]{elsarticle}

\usepackage{amssymb}
\usepackage{amsmath}
\usepackage{booktabs}       % professional-quality tables

\usepackage{subfigure}
\usepackage{subcaption}
\usepackage{comment}

\journal{Nuclear Physics B}

\begin{document}

\begin{frontmatter}



\title{Learning non-ideal vortex flows using the differentiable vortex particle method} %% Article title


\author[1,2]{Ziqi Ji}
\author[1]{Gang Du}
\author[2]{Penghao Duan}
\affiliation[1]{organization={School of Energy and Power Engineering},
            addressline={Beihang University}, 
            city={Beijing},
            postcode={100191}, 
            country={China}}
\affiliation[2]{organization={Department of Mechanical Engineering},
            addressline={City University of Hong Kong}, 
            city={Hong Kong},
            postcode={999077}, 
            country={China}}

%% Abstract
\begin{abstract}
Vortex flows are ubiquitous in both natural processes and engineering applications, including phenomena such as typhoons, water currents, and aerospace fluid dynamics. The vortex particle method, a computational approach grounded in vortex dynamics, has been extensively applied in aerodynamics, oceanography, turbulence, and aeroacoustics. With the recent introduction of machine learning into computational fluid dynamics, a hybrid framework known as the differentiable vortex particle method (DVPM) has been proposed, which integrates the vortex particle method with deep learning to enable efficient learning and prediction. However, the original formulation of DVPM is limited to ideal vortex flow conditions, such as inviscid flows without non-conservative body forces, which significantly restricts its practical applicability. In this study, we extend the differentiable vortex particle method beyond idealized flow scenarios to encompass more realistic, non-ideal conditions, including viscous flow and flow subjected to non-conservative body forces. We establish the Lamb-Oseen vortex as a benchmark case, representing a fundamental viscous vortex flow in fluid mechanics. This selection offers significant analytical advantages, as the Lamb-Oseen vortex possesses an exact analytical solution derived from the Navier-Stokes (NS) equations, thereby providing definitive ground truth data for training and validation purposes. Through rigorous evaluation across a spectrum of Reynolds numbers, we demonstrate that DVPM achieves superior accuracy in modeling the Lamb-Oseen vortex compared to conventional convolutional neural networks (CNNs) and physics-informed neural networks (PINNs). Our results substantiate DVPM's robust capabilities in modeling non-ideal vortex flows, establishing its distinct advantages over contemporary deep learning methodologies in fluid dynamics applications.
\end{abstract}

\begin{graphicalabstract}
\end{graphicalabstract}

\begin{highlights}
\item This study extends the differentiable vortex particle method beyond idealized flow scenarios to encompass more realistic, non-ideal conditions, including viscous flow and flow subjected to non-conservative body forces.
\item We introduce the Lamb-Oseen vortex, an exact solution of the Navier-Stokes equations, as a more suitable training and testing set for our differentiable vortex particle method.
\end{highlights}

%% Keywords

\end{frontmatter}



\section{Introduction}
\label{sec:Introduction}

Vortex flows are pervasive in both natural phenomena and engineering applications, encompassing examples such as typhoons, ocean currents, and aerospace fluid dynamics. Despite their prevalence and critical significance, a comprehensive understanding of vortex flow dynamics remains an open challenge in contemporary scientific research.

The Vortex particle method (VPM) \cite{cottet2001vortex, chorin1993vortex, mimeau_review_2021} is a Lagrangian approach well-suited for simulating vortical flows. It has been widely applied across several fields, including hydrodynamics \cite{bernier_simulations_2019, novati_synchronisation_2017}, aerodynamics \cite{cocle_combining_2008, caprace_wakes_2020}, and porous flows \cite{mimeau_passive_2017, etancelin_improvement_2020}. This method offers several key advantages. First, its Lagrangian nature results in significantly lower numerical dissipation compared to conventional grid-based schemes, allowing for a more accurate representation of wake dynamics. Second, vortex particles move freely, inherently adapting to regions of high vorticity and eliminating the need to resolve regions lacking physical significance. Third, the method's linear structure facilitates the use of acceleration techniques \cite{goude_adaptive_2013, engblom_well-separated_2011, hu_gpu_2013}, which can make VPM extremely computationally efficient.

The concept of hidden fluid mechanics was introduced by Raissi et al. \cite{raissi_hidden_2020} in 2020, referring to fluid mechanical phenomena not explicitly represented by governing equations but instead embedded within flow visualizations. This paradigm aims to decode and learn fluid dynamics directly from visualization data. Currently, there are three principal machine learning approaches to hidden fluid mechanics: the PINNs-based method \cite{raissi_hidden_2020}, the CNN-based method \cite{zhang_learning_2022}, and the DVPM-based approach \cite{deng_learning_2023}. In their pioneering PINNs-based method, Raissi et al. \cite{raissi_hidden_2020} enhanced the conventional PINNs framework by incorporating an additional scalar variable representing marked material concentration to visualize flow patterns. This methodology demonstrated remarkable success in addressing various physical and biomedical applications. Subsequently, Zhang et al. \cite{zhang_learning_2022} developed the CNN-based approach, which integrates physical constraints into the loss function of the CNN framework. Their method effectively utilized particle image velocimetry (PIV) data for flow prediction. More recently, Deng et al. \cite{deng_learning_2023} proposed the DVPM-based method, introducing a deep learning-based vortex particle method specifically optimized for vortex flows. This specialized approach has exhibited superior performance to PINNs and CNN methods in analyzing vortical flow structures.

The DVPM, introduced by Deng et al. \cite{deng_learning_2023}, which integrates deep learning techniques with traditional fluid dynamics approaches. This innovative method harnesses the differentiable properties of neural networks to enhance vortex flow prediction and analysis. However, the method's initial implementation exhibits several limitations that constrain its practical applicability. Specifically, it is restricted to highly simplified vortex flow scenarios, including two-dimensional flows, inviscid conditions, and cases without non-conservative body forces. These constraints significantly limit its extension to more complex, real-world vortex flows. Additionally, a critical consideration arises regarding the method's training dataset, which partially relies on traditional vortex particle method simulations. From the perspective of fluid mechanics researchers, these simulations may not provide sufficient accuracy to serve as reliable ground truth data.

This study extends the differentiable vortex particle method to encompass more general, non-ideal conditions, including viscous flow and flow influenced by non-conservative body forces. As a benchmark case, we introduce the viscous vortex flow, specifically the Lamb-Oseen vortex. The Lamb-Oseen vortex, with its analytical solution derived directly from the NS equations, provides ground truth data for model training and validation. Through rigorous evaluation across a range of Reynolds numbers, we demonstrate that the differentiable vortex particle method achieves superior accuracy compared to CNN-based and PINN-based approaches, as well as the original DVPM formulation. Additionally, we validate our model's capability to accurately predict vortex flow dynamics under the combined influence of viscosity and non-conservative body forces. These results substantiate the robust performance of the DVPM in vortex flow modeling and establish its significant advantages over contemporary deep learning techniques.

The remainder of this paper is organized as follows: Section \ref{sec:Methodology} describes the methodology employed in this work. Section \ref{sec:Results} presents the simulation results. Section \ref{sec:Conclusion} provides the conclusions. \ref{sec:Implementation_detail} provides details of the implementation.

\section{Methodology}
\label{sec:Methodology}

\subsection{Physical model}
\label{sec:physical_model}

For incompressible fluid flows, the vorticity transport equation, derived rigorously from the NS equations, establishes a comprehensive mathematical framework for characterizing the spatiotemporal evolution of vorticity in swirling flow fields:
\begin{equation}
\frac{D \boldsymbol{\omega}}{D t}=\frac{\partial \boldsymbol{\omega}}{\partial t}+\boldsymbol{u} \cdot \nabla \boldsymbol{\omega}=\boldsymbol{\omega} \cdot \nabla \boldsymbol{u}+\nu \nabla^2 \boldsymbol{\omega}+\nabla \times \boldsymbol{b},
\label{eq:vorticity_transport_equation}
\end{equation}
where $\boldsymbol{\omega}$ represents the vorticity vector, $t$ signifies time, $\boldsymbol{u}$ denotes the velocity field, $\boldsymbol{b}$ corresponds to the non-conservative body force, and $\nu$ refers to the kinematic viscosity. For an inviscid, two-dimensional flow in the absence of non-conservative body forces, the equation reduces to $D \boldsymbol{\omega} / D t=\mathbf{0}$, which explicitly demonstrates the Lagrangian conservation of vorticity (i.e., the vorticity of a fluid particle remains constant as it is advected along its trajectory). However, practical flows rarely satisfy these idealized conditions, leading to more complex flow phenomena characterized by intricate vortical dynamics.

Through the application of the Biot-Savart Law, we derive the vortex-induced velocity formulation:
\begin{equation}
\boldsymbol{u}(\boldsymbol{x})=\int K\left(\boldsymbol{x}-\boldsymbol{x}^{\prime}\right) \omega\left(\boldsymbol{x}^{\prime}\right) d \boldsymbol{x}^{\prime},
\label{eq:Biot_Savart_Law}
\end{equation}
where $K\left(\boldsymbol{x}-\boldsymbol{x}^{\prime}\right)$ denotes the kernel function, which is dependent on the displacement vector between the vorticity location and the point of interest in the flow field. In essence, Eq. (\ref{eq:Biot_Savart_Law}) indicates that the velocity at any point within the flow field can be determined by integrating the contributions of vorticity distributed throughout the domain, mediated by the kernel function.

\begin{figure}[htbp]    % h-此处，t-顶部，b-底部，p-单独成页
  \centering
  \includegraphics[width=130mm]{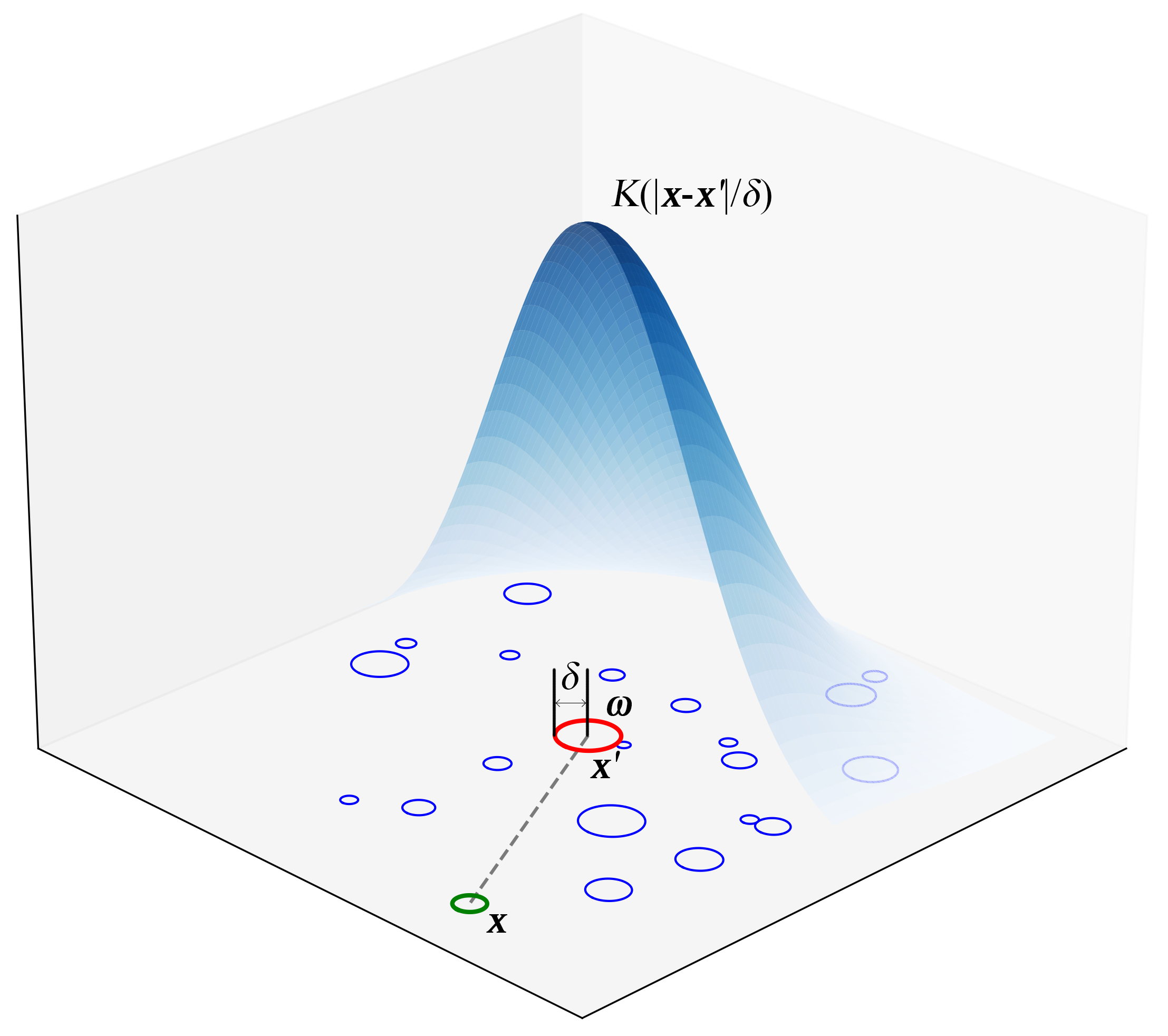}
  \caption{The physical model in our framework. $K\left(\left|\boldsymbol{\boldsymbol{x}}-\boldsymbol{x}\prime\right| / \delta\right)$ represents the kernel function, whose value is determined by multiple parameters: the dependent variable, the Euclidean distance between the vortex core and the point of interest in the flow field $\left|\boldsymbol{x}-\boldsymbol{x}_p\right|$, and the vortex radius $\delta$. Each vortex is characterized by a distinct vorticity $\omega$, which reduces to a scalar quantity in two-dimensional flow configurations.}
  \label{vortex method}
\end{figure}

Fig. \ref{vortex method} presents the physical model employed in our computational framework. Within a two-dimensional flow field comprising $n$ discrete vortices, each vortex is characterized by three fundamental parameters: position $\boldsymbol{x}_i\prime$, vorticity $\omega_i$, and radius $\delta_i$ (assuming circular vortex geometry) for the i-th vortex. These vortices utilize a common kernel function $K\left(\frac{\boldsymbol{x}-\boldsymbol{x}^{\prime}}{\delta}\right)$. Subsequently, the velocity at any arbitrary position $\boldsymbol{x}$ within the flow field can be determined through:
\begin{equation}
\boldsymbol{u}(\boldsymbol{x})=\sum_{i=1}^n K\left(\frac{\left|\boldsymbol{x}-\boldsymbol{x}_i^{\prime}\right|}{\delta_i}\right) \omega\left(\boldsymbol{x}_i^{\prime}\right)
\label{eq:Biot_Savart_Law_discrete}
\end{equation}

By Eq. (\ref{eq:Biot_Savart_Law_discrete}), estimating velocity throughout the flow fields necessitates knowledge of three key attributes of all vortices, along with their kernel function. To address this computational challenge, we implement a framework comprising four neural networks, each dedicated to predicting one of these essential physical quantities. In the two-dimensional flows, the temporal evolution of positional coordinates for n vortices can be expressed as $\left[\left(\boldsymbol{x}_1\right)_t, \ldots,\left(\boldsymbol{x}_n\right)_t\right]$, while their vorticity distributions are represented by $\left[\left(\omega_1\right)_t, \ldots,\left(\omega_n\right)_t\right]$, and their radial parameters are denoted by $\left[\left(\delta_1\right)_t, \ldots,\left(\delta_n\right)_t\right]$. Consequently, we employ four distinct neural networks to characterize these fundamental physical quantities:
\begin{equation}
\left\{\begin{array}{l}
N_{\boldsymbol{x}}(t)=\left[\left(\boldsymbol{x}_1\right)_t, \cdots,\left(\boldsymbol{x}_n\right)_t\right] \\
N_\omega(t)=\left[\left(\omega_1\right)_t, \cdots,\left(\omega_n\right)_t\right] \\
N_{\delta^2}(t)=\left[\left(\delta_{1}^2\right)_t, \cdots,\left(\delta_n^2\right)_t\right] \\
N_K\left(\frac{\left|\boldsymbol{x}_i-\boldsymbol{x}_i^{\prime}\right|}{\delta_i}\right)=K\left(\frac{\left|\boldsymbol{x}_i-\boldsymbol{x}_i^{\prime}\right|}{\delta_i}\right)
\end{array}\right.
\label{eq:NN}
\end{equation}

\subsection{Framework}
\label{sec:Framework}

\begin{figure}[htbp]    % h-此处，t-顶部，b-底部，p-单独成页
  \centering
  \includegraphics[width=130mm]{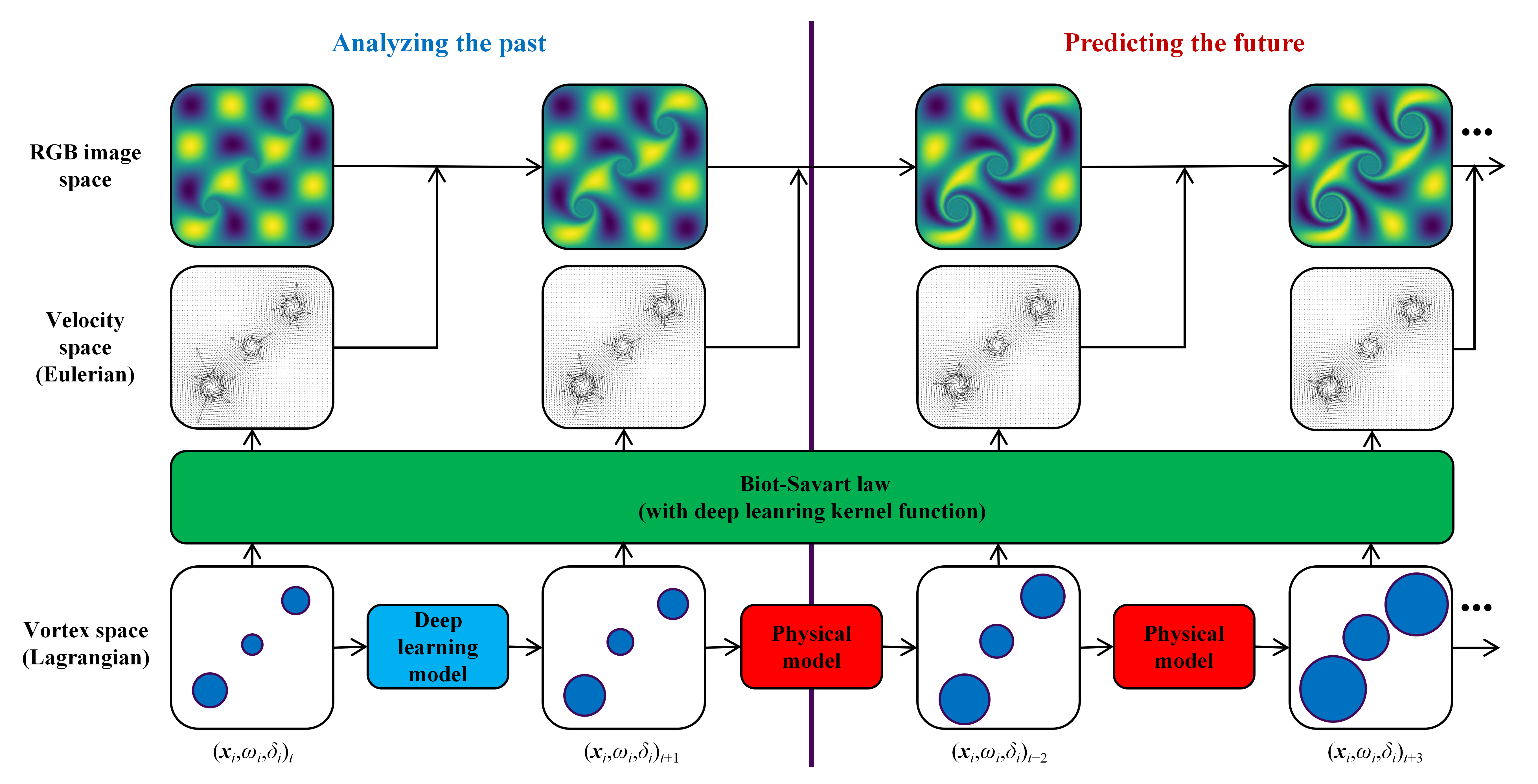}
  \caption{An overview of our framework.}
  \label{flow_chart}
\end{figure}

Fig. \ref{flow_chart} presents a comprehensive overview of our framework. By analyzing historical RGB images, we can extract and track the temporal evolution of vortex attributes (including position, vorticity, and radius) in the Lagrangian frame of reference. We compute the Eulerian velocity field distribution by implementing the Biot-Savart law with an optimized kernel function. Subsequently, we obtain the RGB values for the subsequent time step by advecting the current image data through the calculated velocity field. For future state prediction, while the underlying physical model governs the vortex dynamics, the remaining computational pipeline remains consistent with the past analysis procedure.

We formalize the vortex representation as $V_t=\left[\left(\boldsymbol{x}_1, \omega_1, \delta_1\right), \ldots,\left(\boldsymbol{x}_n, \omega_n, \delta_n\right)\right]_t$. In this framework, $V_t=T(t)$, where $T$ denotes the temporal evolution function of the vortex representation, comprising $N_{\boldsymbol{x}}(t)$, $N_{\omega}(t)$, and $N_{\delta^2}(t)$ as specified in Eq. (\ref{eq:NN}). We employ image-based supervision through the following procedure to optimize the vortex representation $V_t=T(t)$ at time $t$. Initially, we extract a sequence of $m+1$ consecutive frames $\left[I_t, \ldots, I_{t+m}\right]$ from the input video. Subsequently, we derive the velocity field $\boldsymbol{u}_t(\boldsymbol{x})=B\left(V_t\right)$ using Eq. (\ref{eq:Biot_Savart_Law_discrete}), where $B$ represents the Biot-Savart law function incorporating $N_K\left(\frac{\left|\boldsymbol{x}_i-\boldsymbol{x}_i^{\prime}\right|}{\delta_i}\right)$ as defined in Eq. (\ref{eq:NN}). The pair $\left(u_t, I_t\right)$ is then processed through an Eulerian grid integrator to generate the predicted frame $\tilde{I}_{t+1}$. Concurrently, $\left(\boldsymbol{u}_t, V_t\right)$ undergoes integration on Lagrangian particles to yield $\tilde{V}_{t+1}$. This process iterates recursively, utilizing $\tilde{I}_{t+1}$ and $\tilde{V}_{t+1}$ as inputs to generate $\tilde{I}_{t+2}$ and $\tilde{V}_{t+2}$, and so forth. The iteration culminates in a sequence of predicted frames $\left[\tilde{I}_{t+1}, \ldots, \tilde{I}_{t+m}\right]$ originating from time $t$. The optimization of $T$ and $B$ is achieved through end-to-end training by minimizing the discrepancy between the predicted sequence $\left[\tilde{I}_{t+1}, \ldots, \tilde{I}_{t+m}\right]$ and the ground truth sequence $\left[I_{t+1}, \ldots, I_{t+m}\right]$.

Furthermore, we employ a physical model to serve dual purposes: constraining the training process of the temporal evolution function $T$ for vortex representation and predicting the subsequent evolution of vortex flow dynamics. In particular, we establish the differential relationships governing the vortex attributes:
\begin{equation}
\left\{\begin{array}{l}
\frac{d N_{\boldsymbol{x}}(t)}{d t}=\boldsymbol{u}(\boldsymbol{x}) \\
\frac{d N_\omega(t)}{d t}=\nu \nabla^2 \omega+\nabla \times \boldsymbol{b} \\
\frac{d N_{\delta^2}(t)}{d t}=4 \nu
\end{array}\right.,
\label{eq:DN}
\end{equation}
where the temporal derivative $\frac{d N_{\boldsymbol{x}}(t)}{d t}$ is derived from the fundamental differential relationship between displacement and velocity. The expression $\frac{d N_\omega(t)}{d t}$ is formulated based on the vorticity transport equation, as specified in Eq. (\ref{eq:vorticity_transport_equation}), particularly within the framework of two-dimensional flow dynamics. Moreover, the term $\frac{d N_{\delta^2}(t)}{d t}$ is predicated on the decay of vortices due to viscosity, as well as the empirical processes characteristic of the Lamb-Oseen vortex \cite{zuccoli_trapped_2024}. Further implementation details are elaborated in Appendix \ref{sec:Implementation_detail}.

\section{Results}
\label{sec:Results}

In this study, we employ two baseline models for comparison: the PINN-based approach introduced by Raissi et al. \cite{raissi_hidden_2020}, and the CNN-based methodology developed by Zhang et al. \cite{zhang_learning_2022}. We maintain consistent training protocols to ensure a fair comparison across all models. Specifically, rather than using concentration variables as training labels in the PINN-based model, we utilize image RGB values as one of the training labels identical to our proposed model and the CNN-based approach. All images in this study maintain a consistent resolution of 512 $\times$ 512 pixels.

\subsection{Lamb-Oseen vortex}
\label{sec:Lamb-Oseen_vortex}

We selected the well-established Lamb-Oseen vortex to validate our framework. The primary advantage of utilizing the Lamb-Oseen vortex lies in its analytical solution, which serves as a reliable ground truth. This vortex model characterizes the viscous decay of a vortical flow field. The two-dimensional Lamb-Oseen vortex solution is derived from the NS equations in polar coordinates $(r, \theta)$ (representing the radial and tangential directions, respectively), with velocity components $(v_r, v_\theta)$ expressed as:
\begin{equation}
\left\{\begin{array}{l}
v_r=0\\
v_\theta=\frac{\Gamma}{2 \pi r} g(r, t)\\
g(r, t)=1-\mathrm{e}^{-r^2 / 4 \nu t}
\end{array}\right.,
\label{eq:Lamb_Oseen_vortex}
\end{equation}
where $r$ denotes the radial distance from the vortex core and $\Gamma$ represents the circulation strength of the Lamb-Oseen vortex. As demonstrated in Eq. (\ref{eq:Lamb_Oseen_vortex}), the Lamb-Oseen vortex experiences temporal decay due to viscous effects, resulting in a progressive diminution of its strength.

To validate our framework rigorously, we employed the Lamb-Oseen vortex at three distinct Reynolds numbers for training and testing datasets. The Reynolds number, a dimensionless parameter, characterizes the ratio between inertial and viscous forces within a flow. Flows with identical dimensionless parameters exhibit similar characteristics. For the Lamb-Oseen vortex, the Reynolds number (Re) is defined as:
\begin{equation}
Re=\frac{|\Gamma|}{2 \pi \nu}.
\label{eq:Re}
\end{equation}

We employ a passive scalar $s$ to visualize the vortex flows. This passive scalar $s$ satisfies the advection equation presented below:
\begin{equation}
\frac{\partial s}{\partial t}+\boldsymbol{u} \cdot \nabla s=0.
\label{eq:s}
\end{equation}

We establish a square computational domain with dimensions $x \in [-\pi,\pi]$ and $y \in [-\pi,\pi]$, with the Lamb-Oseen vortex core situated at the coordinate origin. The computational domain is discretized using 1024 $\times$ 1024 grid points. According to Eq. (\ref{eq:Lamb_Oseen_vortex}), as $t$ approaches zero, the velocity near the vortex center becomes singular, while the vortex radius diminishes to zero. These characteristics compromise numerical accuracy and represent physical conditions rarely encountered in real-world vortex flows. Consequently, we focus our analysis on the Lamb-Oseen vortex flow at $t = 1$, where the flow exhibits more physically realistic behavior. The passive scalar field is initialized as $s(x, y, t=1) = \sin(x)\sin(y)$. Throughout all simulations, we maintain a constant circulation parameter of $\Gamma = 20 \pi$ while systematically varying the kinematic viscosity as $\nu = 1.0, 0.1, 0.01 \text{ m}^2/\text{s}$ to achieve Reynolds numbers of $Re = 10, 100, 1000$, respectively. Data is sampled at intervals of 0.01 seconds, with each sample constituting a frame in our dataset. Open boundary conditions are implemented on all boundaries of the computational domain.

\begin{figure}[htbp]    % h-此处，t-顶部，b-底部，p-单独成页
  \centering
  \includegraphics[width=130mm]{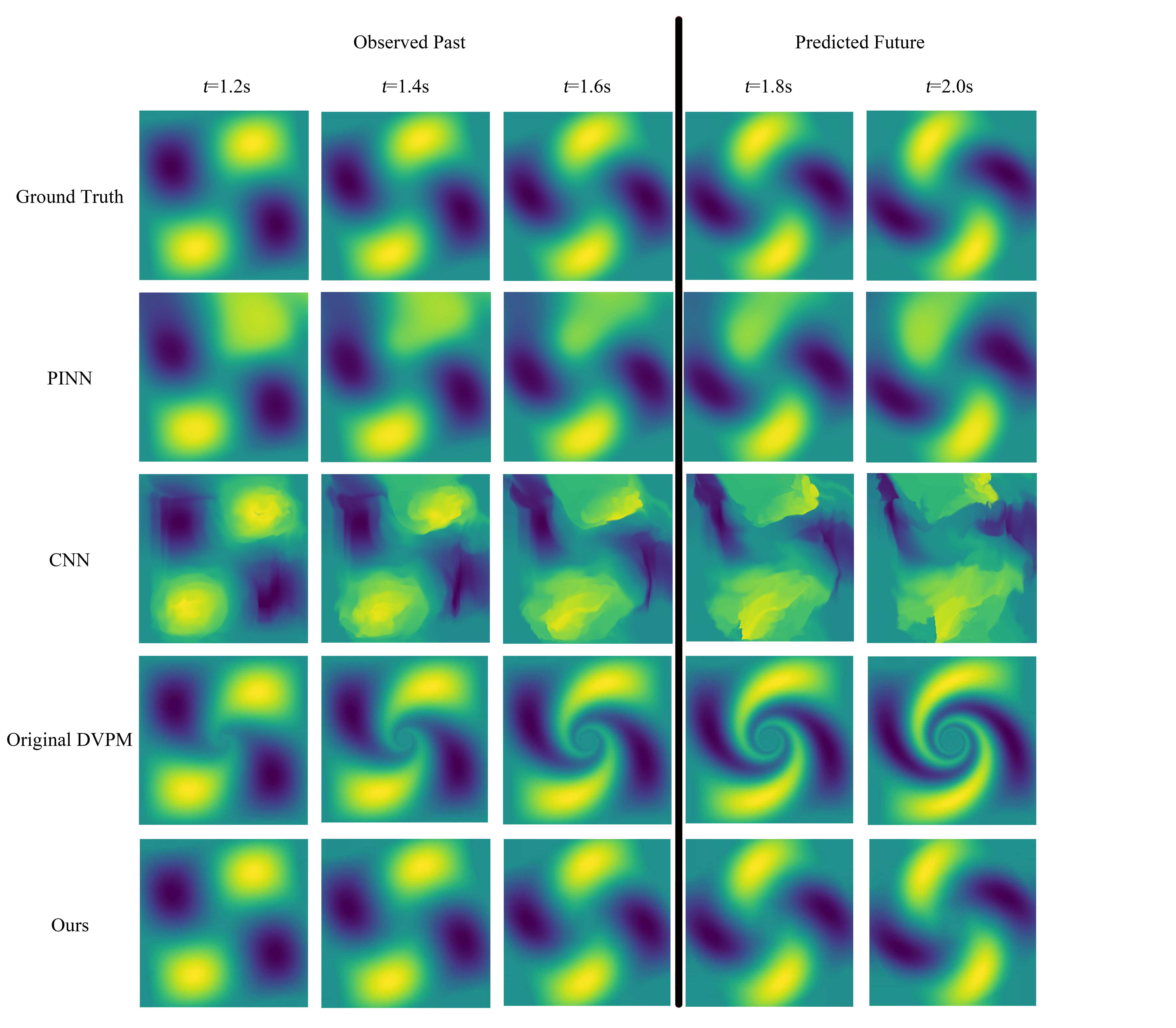}
  \caption{Comparative evaluation of Lamb-Oseen vortex prediction performance at $Re = 10$ among the proposed method and baseline models.}
  \label{lamb_oseen_vortex_Re_10}
\end{figure}

\begin{figure}[htbp]    % h-此处，t-顶部，b-底部，p-单独成页
  \centering
  \includegraphics[width=130mm]{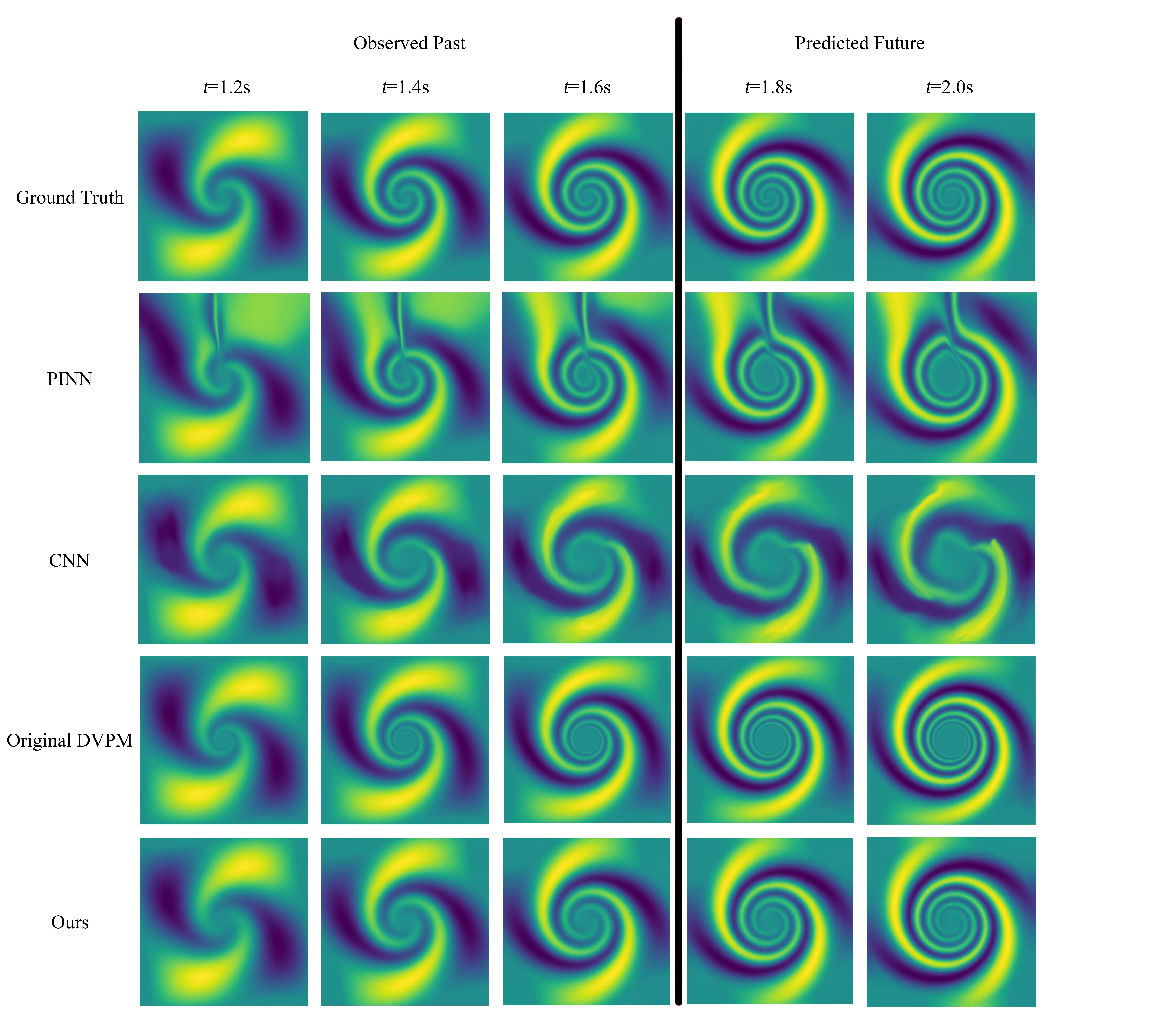}
  \caption{Comparative evaluation of Lamb-Oseen vortex prediction performance at $Re = 100$ among the proposed method and baseline models.}
  \label{lamb_oseen_vortex_Re_100}
\end{figure}

\begin{figure}[htbp]    % h-此处，t-顶部，b-底部，p-单独成页
  \centering
  \includegraphics[width=130mm]{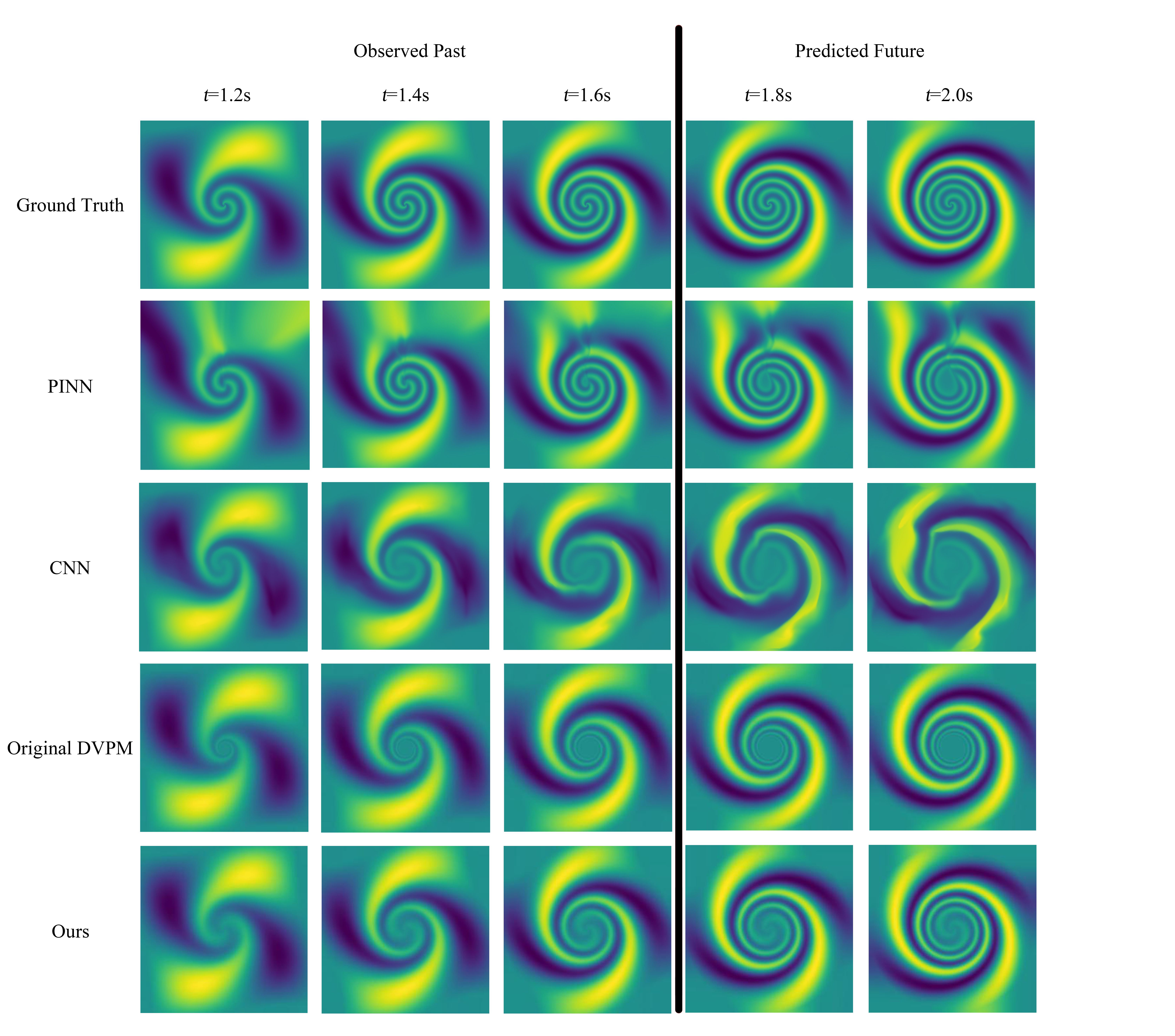}
  \caption{Comparative evaluation of Lamb-Oseen vortex prediction performance at $Re = 1000$ among the proposed method and baseline models.}
  \label{lamb_oseen_vortex_Re_1000}
\end{figure}

Fig. \ref{lamb_oseen_vortex_Re_10}, \ref{lamb_oseen_vortex_Re_100}, and \ref{lamb_oseen_vortex_Re_1000} illustrate the comparative performance of the proposed method against baseline models in predicting Lamb-Oseen vortex dynamics at Reynolds numbers $Re = 10$, $Re = 100$, and $Re = 1000$, respectively. The results demonstrate that our method achieves exceptional accuracy in hindcasting and forecasting scenarios compared to the ground truth data across all examined Reynolds numbers. In contrast, while the PINN approach successfully captures most of the flow regime with reasonable fidelity, it exhibits non-physical patterns in certain regions. The CNN-based predictions, however, manifest significant numerical diffusion and rapidly deteriorate into non-physical flow characteristics as the simulation progresses. The original DVPM demonstrates markedly reduced accuracy for Lamb-Oseen vortices at low Reynolds numbers compared to high Reynolds numbers. This discrepancy arises from the substantially greater influence of viscous effects in low Reynolds number flows, which the original DVPM formulation does not adequately account for.

\begin{figure}[htb]    % h-此处，t-顶部，b-底部，p-单独成页
  \centering
  \includegraphics[width=130mm]{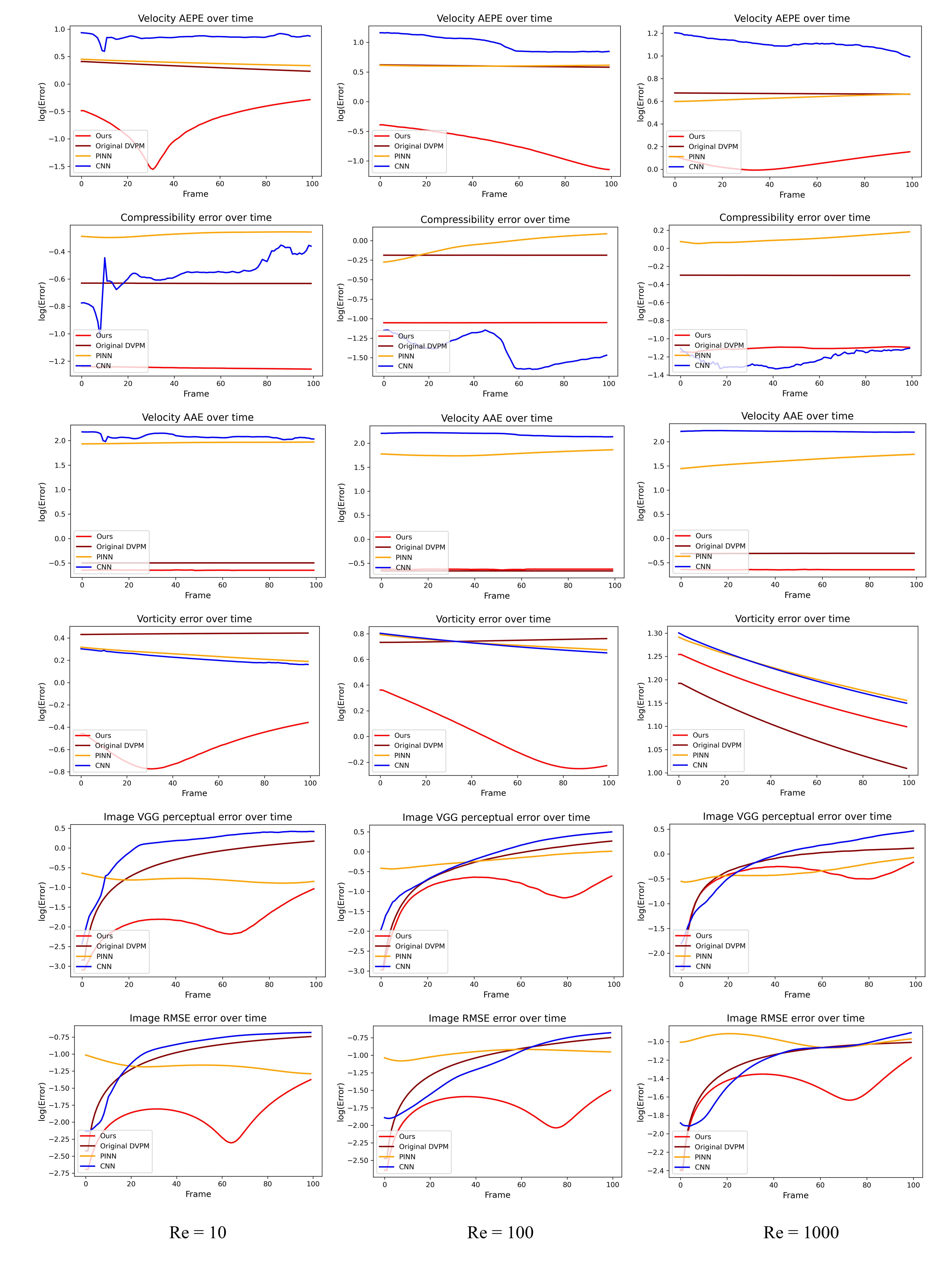}
  \caption{Quantitative assessment of prediction accuracy for the Lamb-Oseen vortical flow field, wherein frames 0-66 comprise the observed historical data sequence, and subsequent frames represent the predicted spatiotemporal evolution of the flow dynamics.}
  \label{Quantitative_error_LO}
\end{figure}

Fig. \ref{Quantitative_error_LO} presents a comprehensive quantitative evaluation of prediction accuracy for the Lamb-Oseen vortical flow field. The analysis encompasses frames 0-66, representing the observed historical data sequence, followed by frames depicting the predicted spatiotemporal evolution of flow dynamics. Six distinct performance metrics are evaluated: velocity average end-point error (AEPE), compressibility error, velocity average angular error (AAE), vorticity error, image VGG perceptual error, and image RMSE error, all assessed over time. Our proposed method demonstrates superior performance across nearly all metrics compared to the baseline approaches.

\subsection{Lamb-Oseen vortex with Coriolis force}
\label{sec:Lamb-Oseen_vortex_with_Coriolis_force}

We employ the Coriolis force as a representative example to evaluate our model's capability in predicting vortex flows subject to non-conservative body forces. We configure the computational domain with an angular velocity of $1 \text{ rad/s}$, oriented perpendicular to the 2D computational domain and directed outward (corresponding to counter-clockwise rotation). The Lamb-Oseen vortex maintains the same configuration with $Re = 1000$ as described in Section \ref{sec:Lamb-Oseen_vortex}.

\begin{figure}[htbp]    % h-此处，t-顶部，b-底部，p-单独成页
  \centering
  \includegraphics[width=65mm]{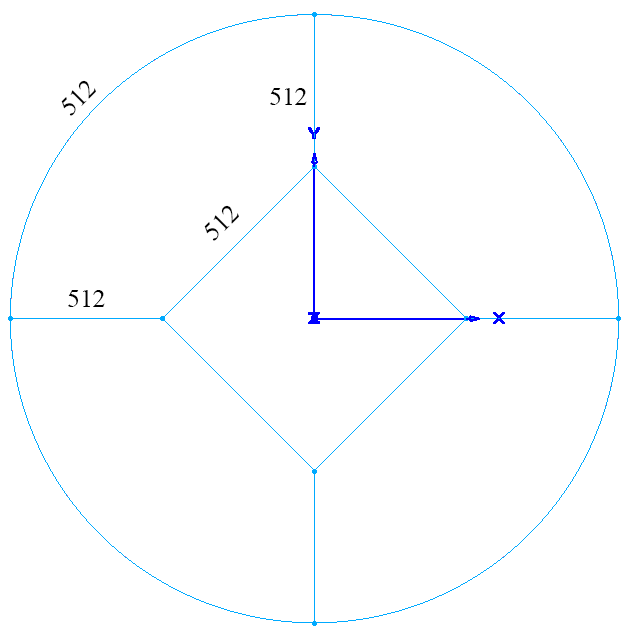}
  \caption{Computational domain configuration and spatial discretizations for the CFD simulation.}
  \label{Mesh_circle}
\end{figure}

Given that a Lamb-Oseen vortex under Coriolis force lacks an exact analytical solution, we employ the CFD software OpenFOAM for numerical simulation. We utilize a circular computational domain for this investigation to mitigate boundary condition effects on both internal and external flows. Fig. \ref{Mesh_circle} illustrates the computational domain configuration and spatial discretization methodology implemented in our CFD simulation. We partition the computational domain into five distinct blocks to enhance mesh quality. The grid point distribution for the upper-left block is explicitly indicated in the figure, while the remaining blocks maintain symmetrical grid point distributions concerning the coordinate origin.

\begin{figure}
\centering \mbox{ \subfigure[]{\includegraphics[width=65mm]{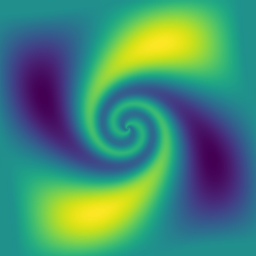}}}\quad \subfigure[]{\includegraphics[width=65mm]{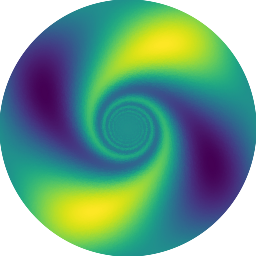}}\quad 
\caption{Comparison of Lamb-Oseen vortex evolution at $t = 1.2 \text{s}$ ($Re = 1000$): (a) without Coriolis effect and (b) with Coriolis force.}\label{lamb_oseen_vortex_Re_1000_comparison}
\end{figure}

Fig. \ref{lamb_oseen_vortex_Re_1000_comparison} illustrates the comparative evolution of the Lamb-Oseen vortex at $t = 1.2 \text{s}$ ($Re = 1000$) with and without the influence of the Coriolis force. The results demonstrate that when the rotational direction of the computational domain aligns with that of the Lamb-Oseen vortex (counter-clockwise rotation), the Coriolis force enhances the vortex structure. Conversely, the vortex is weakened when the rotational directions oppose each other. This observed phenomenon is consistent with the theoretical predictions derived from Eq. (\ref{eq:vorticity_transport_equation}).

\begin{figure}[htbp]    % h-此处，t-顶部，b-底部，p-单独成页
  \centering
  \includegraphics[width=130mm]{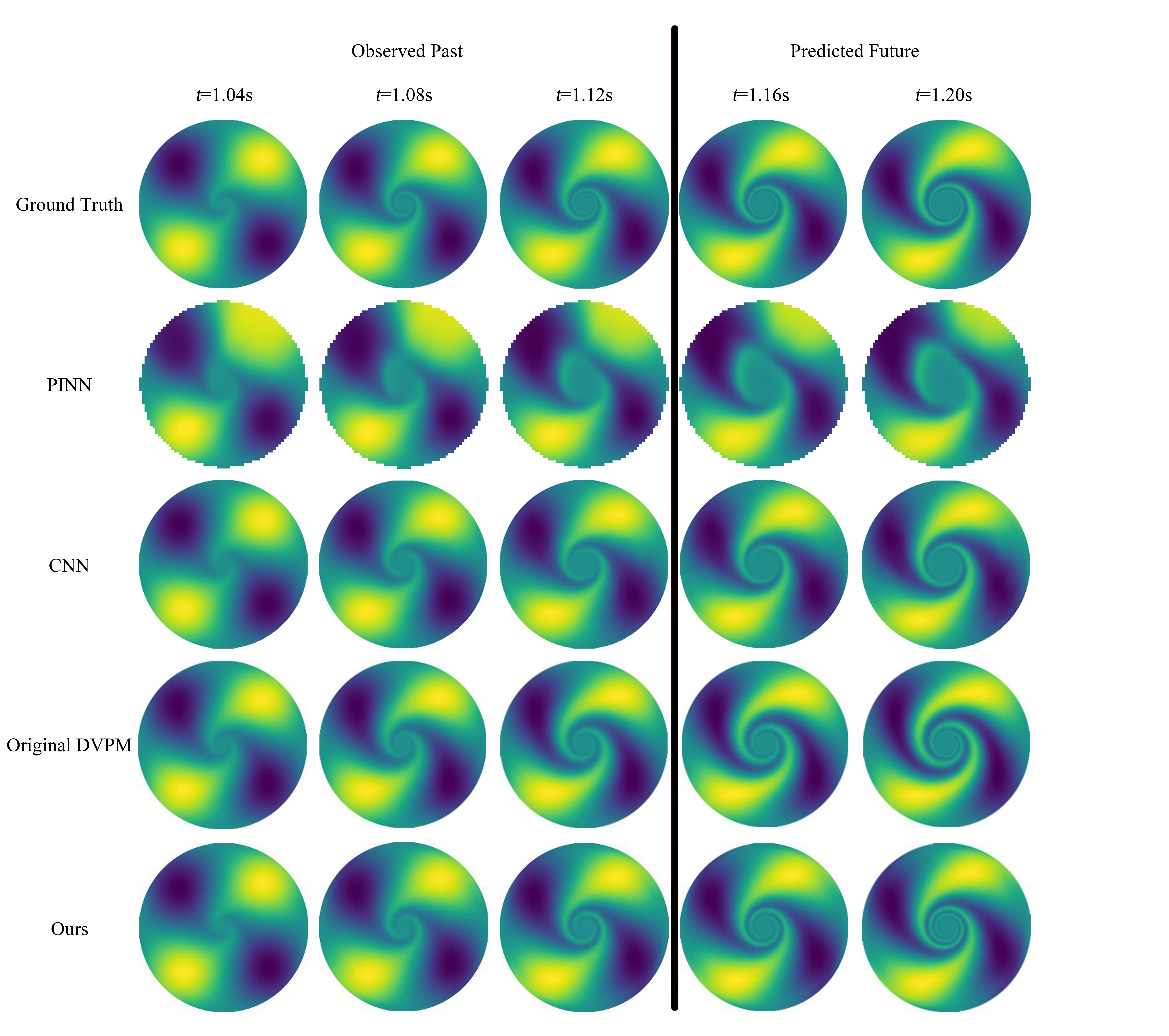}
  \caption{Rigorous comparative assessment of the predictive performance of the proposed methodology against established baseline models for Lamb-Oseen vortex under non-conservative body forces, specifically Coriolis force effects.}
  \label{lamb_oseen_vortex_Re_1000_omega_1}
\end{figure}

Fig. \ref{lamb_oseen_vortex_Re_1000_omega_1} presents a comparative analysis of the prediction results obtained from our proposed model and several baseline models. The results demonstrate that our model accurately predicts the vortex flow dynamics under viscous effects and Coriolis forces. While the CNN-based model successfully captures most of the vortex flow characteristics, it exhibits notable limitations in accurately resolving the core region. In contrast, the PINN-based model yields significantly inferior results compared to the other two approaches, with substantial deviations from the expected flow patterns. The original DVPM predicts a slightly slower rotation compared to the ground truth results.

The quantitative assessment of these vortex flows requires significant effort because CFD simulation results are stored on non-Cartesian grid points of the computational mesh, as illustrated in Fig. \ref{Mesh_circle}. In contrast, deep learning approaches operate on pixel-based representations.

\section{Conclusion}
\label{sec:Conclusion}

In this study, we extend the capabilities of the DVPM to non-ideal conditions, specifically encompassing viscous flows and non-conservative body forces. Furthermore, we utilize the Lamb-Oseen vortex as a more rigorous benchmark for model evaluation, given that it provides an exact solution to the NS equations.

In detail, we employed the Lamb-Oseen vortex at Reynolds numbers of 10, 100, and 1000 as training and test datasets to validate our proposed method. We conducted comprehensive comparisons with PINN-based and CNN-based approaches, as well as the original DVPM. The results demonstrate that our method outperforms these baseline methods across nearly all evaluation metrics. Furthermore, we tested our approach using the Lamb-Oseen vortex with Coriolis force, which also yielded satisfactory results.

Future work will focus on integrating additional physical functionalities derived from traditional VPM research to simulate more complex real engineering flow problems, thereby combining the advantages of both the traditional VPM and deep learning.

\section*{Acknowledgments and Disclosure of Funding}
This research was supported by National Natural Science Foundation of China [No. 52406057], Research Grants Council of Hong Kong [No. CityU 21206123] and National Natural Science Foundation of Guangdong Province [Grant No. 2514050003981].

\section*{Declaration of Generative AI and AI-assisted technologies in the writing process}

During the preparation of this work the authors used GPT-4 in order to improve readability and language. After using this tool, the authors reviewed and edited the content as needed and take full responsibility for the content of the publication.
%%%%%%%%%%%%%%%%%%%%%%%%%%%%%%%%%%%%%%%%%%%%%%%%%%%%%%%%%%%%

\appendix

\section{Implementation detail}
\label{sec:Implementation_detail}

In this section, we describe the implementation details of our proposed method. The networks $N_{\boldsymbol{x}}$, $N_{\omega}$, and $N_{\delta^2}$ are implemented through a sequence of 3 residual blocks with progressively increasing widths of $[64,128,256]$. The network $N_K$ comprises four fully connected layers, each containing 40 neurons. Additional hyperparameters align with those established in previous research \cite{deng_learning_2023}. Prior to the primary training phase, $N_{\boldsymbol{x}}$ undergoes a \text{pretraining} process for 10,000 iterations with two distinct objectives: (1) ensuring that for all $t \in\left[0, t_E\right]$, $N_{\boldsymbol{x}}(t)=\left[\left(\boldsymbol{x}_1\right)_t, \ldots,\left(\boldsymbol{x}_n\right)_t\right]$ precisely corresponds to the centers of an $n \times n$ grid, and (2) establishing that for all $t \in\left[0, t_E\right]$, $\frac{d N_{\boldsymbol{x}}}{d t}=0$, thereby initializing the particles in a stationary state.

Computational performance. Running on a laptop with Nvidia RTX 3090 Ti and AMD EPYC 7R12, our model takes around 0.12 s per training iteration, and around 40000 iterations to converge (for a $256 \times 256$ video with 100 frames). For inference, each advance step costs around 0.01 s.

\bibliographystyle{elsarticle-num}
\bibliography{cas-refs}

\end{document}